\documentclass[preprint,aps,showpacs,floatfix]{revtex4}

\usepackage{graphicx}
\usepackage{amsmath}

\usepackage{soul}

\usepackage{color}

\definecolor{DarkRed}{rgb}{0.45,0,0}
\definecolor{DarkGreen}{rgb}{0,0.40,0}
\definecolor{DarkBlue}{rgb}{0,0,0.45}

\definecolor{LightRed}{rgb}{1.0,0.2,0.2}
\definecolor{LightGreen}{rgb}{0.2,1.0,0.2}
\definecolor{LightBlue}{rgb}{0.2,0.2,1.0}

\definecolor{LightYellow}{rgb}{0.9,0.9,0.6}

\begin{document}

\title{ Aharonov-Bohm effect and plasma oscillations 
in superconducting tubes and rings.}

\author{ E. N. Bogachek }
\author{ I. A. Romanovsky } 
\author{ Uzi Landman }

\affiliation{School of Physics, Georgia Institute of Technology,
             Atlanta, Georgia 30332-0430}


%% 73.20.Mf Collective excitations (including excitons, polarons,
%%       plasmons and other charge-density excitations) (for
%%       collective excitations in quantum Hall effects, see 73.43.Lp)

%% 73.22.-f Electronic structure of nanoscale materials:
%% clusters, nanoparticles, nanotubes, and nanocrystals

%% 74.78.-w Superconducting films and low-dimensional structures
%% 74.78.Na Mesoscopic and nanoscale systems

%% 74.25.-q Properties of type I and type II superconductors
%% 74.25.Ha Magnetic properties

\date{\today}

\begin{abstract}
Low frequency plasma oscillations in superconducting tubes are 
considered.
The emergence of two different dimensionality regimes of plasma 
oscillations in tubes, exhibiting a crossover from
one-dimensional to two-dimensional behavior, depending on whether
$k R\ll 1$ or $k R\gg 1$, where $k$ is the plasmon wave vector
and $R$ is the radius of the tube, is discussed. 
The Aharonov-Bohm effect pertaining to plasma oscillations in 
superconducting tubes and rings, resulting in an oscillatory 
behavior of the plasmon frequency as a function of the magnetic
flux, with a flux quantum period
$hc/2e$ (analog of the Little-Parks effect), is studied. 
The amplitude of the oscillations is proportional to
$(\xi/R)^2$, where $\xi$ is the  superconducting coherence length.
\end{abstract}

\maketitle  

\section{ Introduction}

Collective excitations of charge density, 
so called plasma oscillations, in small low-dimensional
superconducting structures have been a topic of great interest 
for a long time.
There are two types of collective excitations 
in superconductors. One type is the so-called Carlson-Goldman
mode \cite{Carlson_Goldman}.
In this mode the superconducting current oscillations are balanced by
the current of the normal electrons, and the charge densities produced by
the superconducting and normal electrons are mutually compensated.
 This mode occurs only for temperatures that are very
 close to the critical temperature $T_c$ 
\cite{Schmid_Schon,Artemenko_Volkov,CG_ref}.
The other type of collective excitations are plasma oscillations which
are similar to plasma oscillations in normal metals. 
Unlike plasma oscillations in normal metals, those occurring in 
superconductors
cannot exist in bulk superconducting samples because the typical 
frequencies of plasma oscillations in the bulk ($10^{16}$Hz) are far
above the superconducting gap $\Delta$. 
However, in small systems like superconducting wires,
thin films, and tubes, the dispersion relation of plasmons has a
sound-like (acoustic) character.
In what follows we will consider only acoustic type plasma
excitations. We are interested in plasma oscillations that do not
break the Cooper pairs i.e. oscillations with frequencies
$ \omega<2\Delta/{\hbar}$. %\sim {k_B}\sqrt{T_c(T_c-T)}/{\hbar} $ .

The existence of such (acoustic) plasmons in superconductors was
predicted rather early by Kulik~\cite{Kulik}, 
who considered plasma excitations for two geometries: a thin 
infinite solid wire (essentially one-dimensional, 1D) and a
thin infinite film (essentially two-dimensional, 2D). 
It was found that the dispersion relation for
the wire is a linear function of the wave vector $k$ along
the wire, whereas for the infinite thin film the frequency of the 
plasmons is proportional to the square root of the wave vector. 

Formally, plasmons in low dimensional superconductors with
linear and square root dispersion are similar to those in normal
conductors \cite{Ferrel,Bogachek_2,2d-behavior}, but unlike the 
latter they have lower frequencies and decay rates. 
Such ``superconducting plasmons'' were theoretically analyzed
later \cite{Mooij_Schon} and they were subsequently observed 
experimentally in superconducting films
\cite{Buisson_etal,YBaCuO_films} and wires \cite{Caramota_etal}.

The high sensitivity of non-simply connected superconducting
systems (cylinders and rings) to weak magnetic fields is well 
known, and it is manifested by effects such as flux, or fluxoid,  
quantization~\cite{Byers_Yang}, and the Little-Parks effect 
(oscillations of the critical temperature as a function of the 
magnetic flux with a period determined by the superconducting flux 
quantum $hc/2e$~\cite{Little_Parks,Bogachek_3}).
Both of these effects may be considered as manifestations of the 
Aharonov-Bohm (AB) effect~\cite{Aharonov_Bohm} (see e.g. a review
in~\cite{Olariu_Popescu}). Consequently, it is of interest to 
investigate possible manifestations of the AB effect in plasma 
oscillations in superconductors~\cite{note1}.

In this paper we study plasma oscillations in a superconducting
tube with an arbitrary radius $R$ and identify two
dimensionality regimes (1D and 2D) of the plasmon dispersion
relation and the crossover between them, depending on the 
magnitude of $kR$ (see Ref.\cite{Bogachek_1}). 
We demonstrate that the AB effect, pertaining to plasmons in 
superconducting tubes and rings, is expressed as oscillations of 
the plasmon frequencies as a function of the magnetic flux with a 
universal period of $hc/2e$ (the flux quantum of a Cooper pair).

%%% FIGURE 1 WAS HERE

The paper is organized as follows. In Sec. II we discuss the
formalism used to describe plasma oscillations in
superconducting tubes and calculate their dispersion relations. 
In Sec. III we use these results to study the behavior of
plasmons in tubes and rings placed in a
magnetic field. We summarize our results in Sec. IV.

\section{ The dispersion relation for plasma oscillations
 in a hollow superconducting cylinder} 

In this section we study the propagation of plasma waves in a hollow
superconducting cylinder (tube). 
Let us take the symmetry axis of the cylinder as the $z$ axis  
of a cylindrical coordinate system 
and let ${\bf r}=(r,\theta)$, be the radius-vector perpendicular
to the $z$ axis.
For simplicity we assume that the width of the wall of the
cylinder, $d$, is much smaller than its radius $R$ (Fig. 1a).
Since the motion of charge carriers is restricted to be within the
material of the cylinder the charge and current densities can be
written in the form
\begin{eqnarray}
{\bf j}={\bf j}_2 \delta(r-R),\qquad\qquad \rho=\rho_2\delta(r-R),
\end{eqnarray}
where $\delta(r-R)$ is the Dirac delta function and ${\bf j}_2$ and 
${ \rho}_2$ are the two-dimensional (areal) current and charge 
density, respectively. For the sake of brevity we omit the
subscript 2 later.

Both the current flows and the uncompensated charges will
produce electric and magnetic fields around the cylinder.
The Fourier components of the field potentials, $\tilde{\varphi}$
and $\tilde{A}$, and those of the surface charge 
$\tilde\rho$ and the current density 
$\tilde{\bf j}$ {($\tilde{\bf j}=(\tilde j_z,\tilde j_{\theta})$,
where $\tilde j_{z}$ is the component of the current along the $z$
axis, and $\tilde j_{\theta}$ is the circular component)}
corresponding to the frequency ($\omega$), the longitudinal wave 
vector ($k$), and the circular mode number ($m$), satisfy the 
Maxwell equations
\begin{eqnarray}
\frac{1}{r}
\frac{\partial}{\partial r}
\left(r\frac{\partial \tilde{\varphi}}{\partial r}\right)-
\left[\kappa^2+\frac{m^2}{r^2}\right]\tilde{\varphi}=
-4\pi\tilde{\rho }\ \delta (r-R),\\
\frac{1}{r}
\frac{\partial}{\partial r}
\left(r\frac{\partial \tilde{\bf A}}{\partial r}\right)-
\left[\kappa^2+\frac{m^2}{r^2}\right]\tilde{\bf A}=
-\frac{4\pi}{c}\tilde{\bf j }\ \delta (r-R),
\end{eqnarray} 
where $\kappa=\sqrt{k^2-(\omega/c)^2}$ 
is the modified wave vector that takes into account 
retardation effects; here and elsewhere in the paper 
a superscript ``$\sim$'' (for example $\tilde{\varphi}$) denotes a Fourier
transformed quantity.
From the requirement that the potentials  $\tilde{\bf A}$ and
$\tilde{\varphi}$ must be continuous and finite everywhere,
one readily derives the following expressions for the fields 
on the surface 
\begin{eqnarray}
\left\{
\begin{array}{c}
\tilde{\varphi}= 
-4\pi\tilde{\rho} R I_m(\kappa R)K_m(\kappa R),\qquad\qquad\\
\qquad\\
\tilde{\bf A}= 
-\frac{4\pi}{c}\tilde{\bf j} R I_m(\kappa R)K_m(\kappa R).\qquad\\
\end{array}
\right.
\end{eqnarray} 
The electric fields acting on the superconducting and normal 
electrons inside the cylinder can be written as
\begin{eqnarray}
\tilde{E}_z=4\pi i\left( \frac{\omega}{c^2}\tilde{j}^{z} -
k\tilde{\rho} \right)R I_m(\kappa R)K_m(\kappa R), \label{E_z} \\
\tilde{E}_{\theta}=4\pi i\left( \frac{\omega}{c^2}\tilde{j}^{\theta} -
\frac{m}{R}\tilde{\rho} \right)R I_m(\kappa R)K_m(\kappa R), \label{E_phi}
\end{eqnarray}
\noindent
where $I_m(x)$ and $K_m(x)$ are the modified Bessel functions
of the first and second order, respectively.

Our discussion so far was general. 
To describe the superconducting regime we 
adopt a simple two-fluid phenomenological model~{\cite{note2}} with a 
nonlinear superconducting term. We assume that the system is 
almost in a stationary state. For the description of the 
superconducting component we use the time-independent 
Ginzburg-Landau equation. It was shown rather early by 
Bardeen~{\cite{Bardeen}}, that for clean metals (close to $T_c$) 
the two fluid model can be derived directly from the BCS
theory. Subsequently it was also shown that the two-fluid model can 
also be successfully used for dirty metals~{\cite{Rickaysen}}. 
The two-fluid model assumes that electrons are
locally in thermodynamic equilibrium. In the case of low energy
small amplitude long-wavelength collective excitations, like 
acoustic plasma waves, any gradients in the velocities or
densities are sufficiently small, and the equation of motion
of the superconducting electrons takes the form as for classically 
behaving particles.

The electrons of the superconducting component of our
two-fluid model move without dissipation, 
whereas the electrons of the normal fluid dissipate energy.
The parameter that characterizes dissipation of the normal 
fluid $\tau$ is the average time between collisions of the normal
electrons in the metal. Unless this collision time is
very small there is no significant difference between the 
normal and superconducting electrons since both contribute 
to the plasma oscillations. 
The situation is very different, however, for small collision
times $\omega \tau \ll 1$.
In this case the normal carriers almost do not participate in the
plasma oscillations (see discussion in Sec. IV).

Let us assume that the thickness $d$ of the walls of the hollow 
cylinder (tube) is  
smaller than both the superconducting coherence length $\xi$,
and the London penetration depth $\delta $, $(d\ll \delta, \xi)$.
 These assumptions allow us to treat the amplitude of the order
parameter $ \Delta $  in the cylinder as a constant and the 
current density (in the dirty superconductor) can be written 
(from consideration of the time-independent Ginzburg-Landau 
equation) in the form~(see e.g. \cite{De_Gennes})
 \begin{eqnarray}\label{matter}
{\bf j}=eN_s^{eff}{\bf v}+  
                   \sigma_n  {\bf E}, 
\qquad
%\mbox{where}\quad
N_s^{eff}=N_s\left( 1- \frac{v^2}{v^2_c}\right),
\end{eqnarray}
and $N_s$ is the concentration of the superconducting electrons,
${\bf v}=(v_{z},v_{\theta})$ is their velocity which is a 
two component vector: 
$v_{z}$ is the velocity along the tube, $v_{\theta}$ is the 
velocity in the plane perpendicular to the axis of the tube
($v^2=v_{z}^2+v_{\theta}^2$), $v_c$ is the
critical velocity, $\sigma_n$ is the normal conductivity, and 
$E$ is the external electric field. { $N_s^{eff}$ is the 
``effective concentration'' of the superconducting electrons in 
the film, which depends on the velocity $v$.} 
The first term in the equation for the above equation for ${\bf j}$ 
describes the supercurrent, whereas the second term describes the 
current due to the normal electrons and it accounts for all the 
dissipation processes in our system. 

The relationship between the current and the velocity
of the superconducting electrons is nonlinear (see Eq.\ref{matter}). 
When the amplitude of the velocity oscillations in the
plasma wave is much smaller than the critical velocity $v_c$ we
can linearize this relationship about the homogeneous solution ${\bf u}$
\begin{eqnarray}
{\bf v}= {\bf u} +\delta{\bf v},
\end{eqnarray}
where ${\bf u}=(u_z , u_{\theta} ) $ with $u_z={\rm const}$ and 
$u_{\theta}={\rm const}$.
The uniform homogeneous background current through the 
tube has two components: normal and superconducting. 
We assume that the superconductor is very dirty and 
therefore the normal conductance of the tube is very small. 
Because of the very small normal conductance the normal component 
of the constant background current is much smaller than the
total background current and we neglect the voltage drop due 
to the normal current through the tube. In general however, 
if one takes into account the voltage due to the normal current, 
when considering the currents very close to the critical one, 
superconducting states with a uniform time-independent order 
parameter may become unstable toward small perturbations and 
the system may become normal or develop a time-dependent
superconducting state (see 
e.g.~{\cite{Tidecks}}, {\cite{Rubinstein}}). 
We assume here that we are sufficiently far from the critical 
current to assure that such transitions do not occur in the system
under consideration.

Substituting the continuity equation and the equation of motion
written in the Fourier representation 
\begin{eqnarray}
\tilde{\rho}=\frac{m}{\omega R}\tilde{j}^{\theta}+
\frac{k}{\omega}\tilde{j}^z\ ,\qquad
\mbox{ }\qquad
\tilde{\bf v}= -i\frac{e}{\omega m_e}\tilde{\bf E}
\end{eqnarray}

\noindent
into (\ref{E_z}) and (\ref{E_phi}) 
we derive the relations between the currents and the 
velocities of the electrons in the superconductor

\begin{eqnarray}\label{21}
\left(
\begin{array}{c}
\tilde{v}_z \\
\\
\tilde{v}_{\theta}
\end{array}
\right)=
\left(
\begin{array}{cc}
a_{zz} & a_{z \theta}\\
\\
a_{\theta z} & a_{\theta\theta}\\
\end{array}
\right)
\left(
\begin{array}{c}
\tilde{j}_z\\
\\
\tilde{j}_{\theta}
\end{array}
\right)\ ,
\end{eqnarray}

where 

\begin{eqnarray}
a_{zz}=\frac{4\pi e R}{m_e \omega^2}
\left(
k^2-\frac{\omega^2}{c^2}
\right) I_m(\kappa R)K_m(\kappa R)\ ,\\
a_{\theta\theta}=\frac{4\pi e R}{m_e \omega^2}
\left(
\frac{m^2}{R^2}-\frac{\omega^2}{c^2}
\right) I_m(\kappa R)K_m(\kappa R)\ ,\\
a_{\theta z}=a_{z\theta}=\frac{4\pi e k m}{m_e \omega^2}
I_m(\kappa R)K_m(\kappa R)\ .
\end{eqnarray}

After Fourier transformation, the relation between 
the perturbations of the velocity
${ \delta \bf  v}=(\delta v_z , \delta v_{\theta} ) $
and the current density
${\delta \bf j}=(\delta j_z , \delta j_{\theta} ) $ 
can be written as
\begin{eqnarray}\label{27}
\left(
\begin{array}{c}
\delta {\tilde j}_z \\
\\
\delta {\tilde j}_{\theta}
\end{array}
\right)=
\left(
\begin{array}{cc}
b_{zz} & b_{z \theta}\\
\\
b_{\theta z} & b_{\theta\theta}\\
\end{array}
\right)
\left(
\begin{array}{c}
\delta \tilde{v}_z\\
\\
\delta \tilde{v}_{\theta}
\end{array}
\right)\ ,
\end{eqnarray}

\noindent with
\begin{eqnarray}
b_{zz}=N_s ed\left( 1-\frac{u_{\theta}^2+3u_{z}^2}{v^2_c}\right)+
i\omega\sigma_n\frac{m_ed}{e}\ ,\\
b_{\theta\theta}=N_s ed\left( 1-\frac{u_{z}^2+3u_{\theta}^2}{v^2_c}\right)+
i\omega\sigma_n\frac{m_ed}{e}\ ,\\
b_{z\theta}=b_{z\theta} =2N_sed\frac{u_{z}u_{\theta}}{v^2_c}\ .\qquad\qquad 
\end{eqnarray}

Combining equations (\ref{21}) and (\ref{27})
 we get a linear algebraic system of equations
\begin{eqnarray}
\left(
\begin{array}{cc}
a_{zz} & a_{z \theta}\\
\\
a_{\theta z} & a_{\theta\theta}\\
\end{array}
\right)
\left(
\begin{array}{cc}
b_{zz} & b_{z \theta}\\
\\
b_{\theta z} & b_{\theta\theta}\\
\end{array}
\right)
\left(
\begin{array}{c}
\delta\tilde{v}_z\\
\\
\delta\tilde{v}_{\theta}
\end{array}
\right)=
\left(
\begin{array}{c}
\delta\tilde{v}_z\\
\\
\delta\tilde{v}_{\theta}
\end{array}
\right)\  .
\end{eqnarray}
\noindent
In matrix notation the above can be written as
\begin{eqnarray}
\left(
\tilde{\mbox{\bf A}}
\tilde{\mbox{\bf B}}-
\tilde{\mbox{\bf I}}
\right)\delta\tilde{{\bf v}}=0\ ,
\end{eqnarray}

\noindent
where $ \tilde{\mbox{\bf I}} $ is the identity matrix.\\

This system of equations has nontrivial solutions if 
the determinant of the matrix 
$\tilde{\mbox{\bf C}}= \tilde{\mbox{\bf A}}\tilde{\mbox{\bf B}}
-\tilde{\mbox{\bf I}}$ is zero.
The coefficients of matrices ${ \bf \tilde{A}}$ and ${ \bf \tilde{B}}$ are
functions of $k$, $m$ and $\omega$, and the condition
 $\mathcal{D}(k,m,\omega)=
{\rm det}({ \bf \tilde{A}\tilde{B}- \tilde{I}})=0$ 
gives implicitly the desired dispersion relation for the plasma
excitations in the cylinder. Since the resulting expression is
rather cumbersome we do not reproduce it here but represent the result
graphically in Fig.\ref{om_vs_kz_cylinder3} (for $\sigma_n=0$).

%%% FIGURE 2 WAS HERE

Assuming that the normal conductivity of the 
material of the superconductor is zero, the general relation 
for the frequency of the plasmons can be written approximately 
 in the form 
\begin{widetext}
\begin{eqnarray}\label{tube_dispersion}
&&\omega^2=
\omega_s^2\left(\frac{d}{R}\right)
\left[
k^2R^2\left(1-\frac{3u_z^2+u_{\theta}^2}{v_c^2}\right)+
\right. \\
&&\qquad\qquad\qquad\qquad
+\left.
m^2\left(1-\frac{u_z^2+3u_{\theta}^2}{v_c^2}\right)-
4mkR\frac{u_zu_{\theta}}{v_c^2}
\right]I_m(\kappa R)K_m(\kappa R)\ , \nonumber
\end{eqnarray}

\end{widetext}

\noindent
where $\omega_s=\omega_0\sqrt{N_s/N}$ \ , \  $N$ is total
concentration of electrons (normal and superconducting) and
$\omega_0=\sqrt{{4\pi e^2 N}/{m_e}}$ is a frequency of plasma
oscillations in a bulk normal metal. In deriving 
Eq.\ref{tube_dispersion} we neglected terms of the order of
$\omega_s^2 Rd/c^2 \ll 1$, which are related to
relativistic retardation effects and are small for tubes
of practical sizes.

There are two important cases for which the dispersion relation
for the plasma oscillations can be written explicitly in a simple form.
These are: (i) the limiting case of plasma oscillations in an infinite
superconducting thin wire (1D case), and (ii) plasma oscillations
in a thin superconducting thin film (2D case).

In the first case (i), the radius $R$ should be
reduced to the limit when the cylinder becomes a thin
thread without a hole inside. In this limit ($R=d$)
 one gets for the circular mode with $m=0$, 
when $u_{\theta}=0$
\begin{eqnarray}\label{limit1}
\omega=\omega_s k d
\sqrt{
  \left(
  1-\frac{3u^2}{v^2_c}\right)
  \ln\left( \frac{2}{\gamma k d}  
  \right)
},
\end{eqnarray}
where we used asymptotic expressions for the modified Bessel 
functions for small values of the arguments, $x\ll 1$ 
($I_0(x)\to 1$ and $K_0(x)\to -{\rm ln}(\gamma x/2)$),
and the constant $\gamma=exp(C)\approx 1.781$ is the exponent of Euler's
constant. Such a linear dispersion relation is typical for
one-dimensional conductors.

To obtain the dispersion relation for the thin superconducting
thin film (case(ii)) one should take the large radius limit for the cylinder
($k R\gg 1$). In this case we obtain a square root dispersion relation
($u_{\theta}=0$)
\begin{eqnarray}\label{limit2a}
\omega^2=\frac{\omega^2_s k d}{2}
\left(1-\frac{3u^2}{v^2_c}\right)\ .
\end{eqnarray}
Note that Eqs. (\ref{limit1}) and (\ref{limit2a}) reproduce
the expressions derived earlier by Kulik  
for superconducting thin wires and films 
( see Eqs. (14) and (17) in Ref.~\cite{Kulik}, where the thin
wire is referred to as a ``filament'').
Similarly, using asymptotic expressions for the modified Bessel
functions of high orders (i.e. $m\gg 1$) we 
obtain for $kR\to 0$ the expression ($u_{\theta}=0$)
\begin{eqnarray}\label{limit2b}
\omega^2=\frac{\omega^2_s m d}{2 R}
\left(1-\frac{u^2}{v^2_c}\right)\ .
\end{eqnarray}
This expression shows that for large $m$ the plasmon frequency 
$\omega$  is proportional to the $\sqrt{m}$, where $m$ is the 
circular mode number.

An interesting property that emerges from the expression given in 
Eq. (\ref{tube_dispersion}) is that the frequency of the plasma 
oscillation can be decreased by passing an electric
current through the tube. By increasing the current 
(i.e. increasing the velocities $u_z$ and $u_{\theta}$ in 
Eq. (\ref{tube_dispersion})) and bringing it
close to the critical current in the film, $j_{max}$,
one can lower the frequency of the plasmons and cause it to take 
values that lie below the energy gap. 
This lowering can be achieved for a range of wave vectors 
that is large enough to allow observation of a dimensionality 
crossover from a 1D ($k R\ll 1$) to a 2D ($k R\gg 1$) 
behavior \cite{Bogachek_1};
{ see Fig. {\ref{om_vs_kz_cylinder3}}
where we display the plasmon frequency 
as a function of the dimensionless radius $kR$, for several
values of the current $I$ (expressed in terms of $\eta=I/I_c$)
through the tube ($I_1<I_2<I_3$).
Note that higher currents through the tube correspond to lower
frequencies of the plasmons. The observed decrease of the
plasmon frequency originates from the fact that, 
according to Eq. ({\ref{matter}}), increasing the current along 
the tube involves an increase in the velocities (i.e.  $u_z$)
of the superconducting electrons in the tube which in turn causes
an effective decrease in the concentration of the superconducting 
electrons that participate in plasma oscillations
and determine the plasma frequency.
}

\section{Superconducting tubes and rings in an external magnetic field}
In this section we analyze the influence of a weak magnetic
field on the propagation of plasma excitations in superconducting
microstructures. We consider plasmons in two geometries: tubes
and rings.

\subsection{Tubes}
Let us consider the situation when the superconducting microcylinder is 
placed in a longitudinal magnetic field $H$ (Fig. 1a). We assume that
the magnetic field is parallel to the symmetry axis of the cylinder and
that it is weak enough such that the system remains
superconducting.  This geometry allows observation of several
interesting effects, such as quantization of the magnetic flux through a
hole in the cylinder, and periodic dependence of the critical temperature.
One may also inquire about the influence of the magnetic field
on the dispersion relation of the plasma oscillations. 

Since the cylinder wall is made from a very thin film
(we take the width of the wall of the cylinder to be 
smaller than both the London penetration length $\lambda$, the
coherence length $\xi$, and the magnetic length 
$l_{H}=\sqrt{\Phi_0/H}$), the amplitude of the superconducting
order parameter $\vert\Psi\vert$ is constant across the wall of 
the cylinder. 
The magnetic field penetrates into the wall and the 
flux is not quantized. The quantity that is quantized in this
case is the total change of the phase of the order parameter (so 
called 'fluxoid')~\cite{De_Gennes}. Due to quantization of the 
fluxoid, the average circular velocity $u_\theta$ of the
electrons in the thin-walled cylinder is a periodic function of 
the magnetic flux
\begin{eqnarray}
u_\theta=\min\limits_n
\left[
\frac{\hbar}{m_e R}
\left( 
n-\frac{\Phi}{\Phi_0} 
\right)
\right]\ ,
\end{eqnarray}
\noindent
where $\Phi=\pi R^2 H$ is the magnetic flux through the cylinder and
$\Phi_0={\pi\hbar c}/{e}$ is the superconducting quantum of the 
magnetic flux. { The notation $\min\limits_n [\dots] $ in the
above formula denotes that for a given value of the flux $\Phi$ one 
should take a value of $n$, which minimizes the velocity $u_{\theta}$}.
One should remember, however, that in order to observe the above
quantization phenomenon the radius of the cylinder should be
not too small, in order to allow that even weak fields, that do not
destroy the superconductivity, may create fluxes $\Phi$ of the order
of the flux quantum. Note that since the wall thickness $d$ is small,
Little-Parks oscillations ($\pi R^2 H > \Phi_0$) 
might be observed in samples with $\pi R^2 H_c < \Phi_0$,
where $H_c$ is the critical magnetic field for the bulk
{~\cite{liu_zadorozhny_etal}}.
%
%%% FIGURE 3 WAS HERE
%
Inserting the above expression for the velocity $u_{\theta}$ into 
Eq.(\ref{tube_dispersion}) and considering for simplicity only 
the case with $m=0$, we obtain
\begin{equation*}\label{tube-in-mfield}
\omega^2=
\frac{4\pi e^2 N_s^{eff}}{m_e} k^2 R d ~ 
I_0(\kappa R)K_0(\kappa R)\ , \tag{25a}
\end{equation*}
where
\begin{equation*}\label{tube-in-mfield_b}
N_s^{eff}=N_s
\left( 1-3\frac{u_z^2}{v_c^2}-\frac{\hbar^2}{m_e^2 R^2 v_c^2}
\min\limits_n
\left[
\left(
n-\frac{\Phi}{\Phi_0}
\right)^2
\right]
\right) \ . \tag{25b}
\end{equation*}
From Eq.({25}) we can conclude that the
frequencies $\omega$ of the plasmons, as
well as their velocities, demonstrate an Aharonov-Bohm (AB) 
behavior i.e. a periodic dependence on the magnetic flux with the
fundamental period $\Phi_0$ (Fig. \ref{tube-plasmons}). 
The amplitude of the oscillations of the frequency 
for a cylinder with a large radius ($R > \xi $) is
approximately proportional to $(\xi/R)^2$, where $\xi$ is
the coherence length of the superconductor.

{
To illustrate the above analysis, we show in 
Fig. {\ref{tube-plasmons}} the behavior of the plasmon
frequencies, $\omega$, in a superconducting tube as a function 
of the dimensionless flux $\Phi/\Phi_0$. 
In Fig. {\ref{tube-plasmons}}(a)  we display $\omega$ for 
various values of the current,
characterized by the parameter $\eta=I/I_c$ where $I_c$ is the 
critical current, for a fixed temperature, 
and in Fig.~{\ref{tube-plasmons}}(b) we show $\omega$ for 
several temperatures, while keeping the current at a constant
value $I=0.1 I_c$.
The numerical results presented in Fig.~{\ref{tube-plasmons}} 
were obtained from the transcendental equation
({25})
(where the arguments of the modified Bessel functions depend on 
$\kappa=\sqrt{k_z^2-\omega^2/c^2}$).
Note (see Fig. {\ref{tube-plasmons}}(a)) that the amplitudes of 
the AB oscillations of the frequencies are larger for higher
currents through the tube, while at the same time the absolute 
values of the frequencies decrease for larger currents. 
The latter decrease of the plasmon frequency for higher currents 
was discussed in the context of Fig. {\ref{om_vs_kz_cylinder3}} 
at the end of Sec. II.
The increase of the amplitude of the oscillation of the plasmon 
frequencies for larger currents is due to the nonlinear
dependence of the plasmon frequency on the effective 
concentration of superconducting electrons 
($\omega\sim\sqrt{N_s^{eff}}$) in Eq. ({25}). 
As a result of this nonlinearity, for a lower 
concentration of the superconducting electrons (caused by the
flow of current, or due to higher temperature, with either of 
these resulting in effective lowering of the concentration of
electrons which may participate in collective plasma
oscillations) a given change in the concentration due to 
magnetically induced circular currents will yield a relatively 
larger variation of the plasma frequency.
The decrease of the plasmon frequency as the temperature is 
increased is illustrated in Fig. {\ref{tube-plasmons}}(b). 
}

\subsection{Rings}

We consider here a superconducting ring made from a wire
of diameter $d$, with the radius of the ring, $R$, being much 
larger than $d$ (i.e. $R\gg d$, see Fig.1b). 
For small currents in the ring we can neglect 
interactions between different parts of the wire and consider the 
ring  as a straight superconducting wire with imposed periodic 
boundary conditions. Let  us fix a point on the ring as the
origin of a local coordinate system and let $x$ be the coordinate 
along the wire and ${\bf r}$  the coordinate perpendicular 
to the wire. The relation between the Fourier components of
small perturbations of the charge density and the electrostatic 
potential can be written as
\begin{eqnarray}
\delta\tilde{\phi}(\nu,\omega)= 
\delta\tilde{\rho}_{1D}(\nu,\omega) 
\ln\left( \frac{\nu^2+\nu_c^2}{\nu^2}\right) \ ,
\end{eqnarray}
\noindent where $\nu=1,2,3,\dots $ is now the discrete
dimensionless wave number of the plasma oscillations along the ring 
(which is related to the quantized wave vector $k$ 
along the ring as $\nu=k R$), and $\nu_c$ 
is the cut-off parameter, $\nu_c\sim R/d$, which arises
because of the finite diameter of the wire from which the ring 
is made ($\nu \ll \nu_c$). Using the relation between
the scalar potential and the electric field 
$\tilde{E}(\nu,\omega)=i\nu\tilde{\phi}(\nu,\omega)/R$ ,
the equation of motion 
$\omega \tilde{v}(\nu,\omega) = -ie\tilde{E}(\nu,\omega)/m_e$
and the continuity equation 
$k\tilde{I}(\nu,\omega) =\omega \tilde{\rho}_{1D}(\nu,\omega) R$
written in the Fourier representation, we find the expression
that connects the perturbation of the carrier velocity with 
the perturbation of the current 
\begin{eqnarray}\label{35}
\delta \tilde{v}(\nu,\omega)=
\delta \tilde{I}(\nu,\omega) 
\frac{e \nu^2}{\omega^2 m_e R^2}
\ln\left( \frac{\nu^2+\nu_{c}^2}{\nu^2}\right)\ .
\end{eqnarray}
At this point we use again a linearized form of Eq. (\ref{matter}) 
\begin{eqnarray}\label{36}
\delta I(\nu,\omega)=
eN_sS\left(1-\frac{3u^2}{v^2_c}\right)\delta v(\nu,\omega) \ ,
%+i\omega\frac{\sigma_n S m_e}{e}\delta v
\end{eqnarray}

\noindent where $S=\pi d^2/4$ is the cross section of the
wire making up the ring. 

Following the same arguments as those used by us in our discussion of
the cylindrical system under the influence of a magnetic field,
we can write again the expression for the uniform component of
the velocity as
\begin{eqnarray}
v_0=
\min\limits_n
\left[
\frac{\hbar}{m_e R}
\left( 
n-\frac{\Phi}{\Phi_0} 
\right)
\right]\ .
\end{eqnarray}

%% where $\Phi$ is the magnetic flux through the ring and
%%  $\Phi_0={\pi\hbar c}/{e}$ is the quantum of the magnetic flux

Combining this result with Eqs. (\ref{35}) and (\ref{36}) we
arrive at the dispersion relation for plasma 
oscillations in the ring (Fig. \ref{ring-plasmons})
\begin{eqnarray}\label{ring_dispersion}
\omega^2=\frac{N_se^2\nu^2S}{m_eR^2}
\left(
1-\frac{3\hbar^2}{m_e^2R^2v^2_c}
\min\limits_n
\left[
\left(n-\frac{\Phi}{\Phi_0}\right)^2
\right] 
\right) 
\ln\left(
\frac{\nu^2+\nu^2_c}{\nu^2}
\right)\ .
\end{eqnarray}

%%% FIGURE 4 WAS HERE

For small wave numbers $\nu\ll \nu_c$ 
the frequency of the plasma oscillation $\omega$ can be 
approximated by
\begin{equation}\label{ring-disp-f1}
\omega\sim (\nu/R) c(\nu)\ , 
\end{equation}
where $c(\nu)$ is the velocity of the plasmons showing an AB
behavior, expressed as a periodic function of the magnetic flux, 
\begin{equation}
  c(\nu)=\sqrt{\frac{2N_se^2S}{m_e}
\left(
1-\frac{3\hbar^2}{m_e^2R^2v^2_c}
\min\limits_n\left[
\left(
n-\frac{\Phi}{\Phi_0}
\right)^2
\right]
\right) 
\ln\left(\frac{\nu_c}{\nu}\right)} \ . 
\end{equation}
The oscillation amplitude of the plasmon frequency in 
large rings $R > \xi$ is proportional to $(\xi/R)^2$, 
which is similar to the case of the plasma oscillations in a
tube (see Sec. III.A).

{
Characteristic properties of the plasma oscillation frequencies 
in superconducting rings are illustrated in 
Figs. {\ref{ring-plasmons}}(a) and {\ref{ring-plasmons}}(b).
In Fig. {\ref{ring-plasmons}}(a) we display the plasmon
frequencies as
a function of the mode number $\nu$ for different values of
the dimensionless flux, $\Phi/\Phi_0$; 
the mode numbers are discrete because of the periodic boundary 
conditions in the ring. 
In accordance with Eq. ({\ref{ring-disp-f1}}) higher modes 
(larger $\nu$) correspond to higher frequencies
~{\cite{note3}}.
Applied magnetic flux through the ring induces a circular
persistent current, which reduces the 
effective concentration of superconducting electrons 
participating in the plasma oscillations, with a consequent
lowering of the plasmon frequency.
Since the induced current is a periodic function of the magnetic
flux, the frequency of the plasma oscillations for each mode is
also a periodic function of the magnetic flux with a period 
$hc/2e$ (see Fig.{\ref{ring-plasmons}}(b)).
}
 
\section{Summary.}

In this paper we have studied collective charge density
 oscillations (plasmons) in superconducting microtubes and microrings. 
Using a simple two-fluid model for the superconductor, we derived 
the dispersion relation for plasmons in a cylindrical tube of
radius $R$, i.e. the plasmon frequency $\omega$ as a function of
$k R$. %(\ref{tube_dispersion}).
We have demonstrated that depending on the magnitude of $k R$, a
 crossover emerges where the plasmon dispersion relation
 changes from a linear dependence on $k R$ 
(Eq. (\ref{limit1}), the 1D limit)
to a square root dependence (Eq. (\ref{limit2a}, the 2D limit). 
The behavior in these limiting cases 
%of Eqs. (\ref{limit1}) and (\ref{limit2a}) 
is in agreement with previous theoretical predictions 
\cite{Kulik,Mooij_Schon} and experimental observations 
\cite{Buisson_etal,YBaCuO_films,Caramota_etal}.

We have also considered the effects of weak magnetic fields
on charge density excitations in
superconducting microtubes and microrings, and we have shown 
that the dispersion relations for the plasmons are oscillatory 
functions of the magnetic flux with a universal period of 
$hc/2e$, and an amplitude of the order of $(\xi/R)^2$.
Such behavior of the plasmons in superconducting microstructures 
is a manifestation of the Aharonov-Bohm effect.

In conclusion we discuss briefly dissipation effects in our 
systems.
Our model does not take into account dissipations due to
variations of the order parameter. We considered very long 
wavelength plasma oscillations characterized by small 
amplitudes $\vert\delta \Delta \vert \ll \vert \Delta\vert$, 
and frequencies that are restricted by the superconducting gap 
frequency ($2\Delta/\hbar$). For these conditions dissipation
due to variations of the order parameter 
$\sim \vert \partial \Delta /\partial t \vert^2$ is
negligibly small.

The expressions given in (\ref{tube_dispersion}) and 
(\ref{ring_dispersion})  for the  
dispersion relations of plasmons in superconducting tubes and 
rings are written for cases where the normal conductivity 
of the superconductors can be neglected. 
In general, a nonzero normal conductivity $\sigma_n$ adds 
to the dispersion relation an imaginary term which expresses 
energy dissipation and decay of the plasma oscillations.
For example, for superconducting tubes one can 
write ($\bar\omega=\omega(\sigma_n=0)$)
\begin{eqnarray}
  \omega(\sigma_n)\approx\bar\omega+i\gamma, 
\end{eqnarray}
where for the mode with $m=0$ the ratio of the imaginary part 
of the frequency to the real part is 
\begin{eqnarray}\label{decay-ratio}
\gamma/\bar\omega\sim 
({\sigma_n}/{\omega_s}) k\Bigl[Rd~I_0(kR) K_0(kR)\Bigr]^{1/2}.
\end{eqnarray}
For the limiting case of a thin wire (1D), $kR\ll 1$, 
Eq. (\ref{decay-ratio}) gives 
$\gamma/\bar\omega\sim ({\sigma_n}/{\omega_s}) k
\sqrt{Rd~\ln(1/(kR))}$. In the case of a thin film (2D), 
$kR\gg 1$, damping effects are described by the expression 
$\gamma/\bar\omega\sim ({\sigma_n}/{\omega_s}) \sqrt{kd}$, in
agreement with the results of Ref.~\cite{Kulik}.

The plasmon damping, given by Eq. (\ref{decay-ratio}) 
is small for dirty superconductors ($\sigma_n\to 0$).
Higher normal conductivity results in stronger dissipation and 
decay of the plasma oscillations. This counterintuitive result can be 
explained qualitatively by the following arguments. 
When the collision time is small the coherent motion of the
plasma waves is created by the superconducting electrons, whereas 
normal electrons
are only partly involved in this motion, and the faster they 
achieve equilibrium by collisions the better they follow the
collective motion of the other electrons in the plasma wave, 
and as a result the system 
evolves more adiabatically with less dissipation.\\

Acknowledgement: This study was supported by the U.S. Department
of Energy, Grant No. FG05-86ER45234.

\vspace{1cm}

\newpage
\begin{center}
{\large FIGURE CAPTIONS}
\end{center}

{FIG. 1:}
{
  A schematic of a superconducting cylinder and a ring. 
  The width, $d$, of the cylindrical shell (a) and the diameter, 
  $d$, of the wire from which the ring (b) is made of, are taken 
  to be small, i.e. $d\ll R$ and $d\ll \xi$, where $\xi$ is 
  the correlation length of the superconductor.}

\vspace{2mm}

{FIG. 2:}
{The dispersion relation ($\hbar\omega/\Delta$\ \  versus \  $kR$) 
  for longitudinal plasmons in a superconducting cylinder. 
   Here $\Delta$ is the energy gap for electron-hole-like
    excitations inside the superconductor, for which we use 
  an estimate (near $T_c$) 
  $\Delta\sim {k_B}\sqrt{T_c(T_c-T)}$~{\cite{De_Gennes}}.
  For long wavelength
  ($k R < 1$, see inset) plasmons, the dispersion relation is 
  approximately 
  linear, whereas for plasmons with a shorter wavelength, 
  $k R > 1$, the dispersion relation is
  similar to the one for plasmons in a thin superconducting
  film ($\omega \sim \sqrt{k}$). The three different curves in 
  the figure correspond to three different 
  currents through the tube with $I_1< I_2 < I_3$, with the   
  current $I$ expressed in terms of the critical current, 
  $\eta=I/I_c$.
  {For each value of the current, we have found from 
  Eq. ({\ref{matter}}) the corresponding value of the uniform
  background velocity $u_{\theta}$ and then substituted it in 
  Eq. ({\ref{tube_dispersion}}). Since higher velocities of the 
  superconducting electrons correspond to lower effective 
  concentrations $N_s^{eff}$}, 
  the frequency of the plasma oscillations is lower for the
  larger currents through the tube.  
}

\vspace{2mm}

{FIG. 3:}
{
Aharonov-Bohm oscillations ($\omega$
vs the dimensionless flux, $\Phi/\Phi_0$) of the dispersion
relation for plasmons in a thin superconducting cylindrical tube.
The parameters of the hollow tube are the following: radius 
$R=5\times 10^{-4}cm$, width of the wall $d=10^{-7}cm$;
The tube is made from niobium ($T_{c}=9.3K$),
and for the order parameter $\Delta$ we use the same 
estimate as in Fig. 2.
Both figures correspond to plasmons in the zero circular mode 
$m=0$ and a longitudinal wave vector $k=0.005/R$.
In {\bf (a)} we show the frequency $\omega$ 
of the plasma oscillations as a function of the dimensionless
magnetic flux for a fixed temperature $T=8.5K$, for different 
values of the parameter $\eta=I/I_c$, where $I$ is the 
current along the tube, and $I_c$ is the critical current.
In {\bf (b)} we show for a fixed current along the tube
($I=0.1 I_c$) the frequency of 
the plasma oscillations as a function of the dimensionless 
magnetic flux, for a range of temperatures.
}

\vspace{2mm}

{FIG. 4:}
{
Frequencies of plasma oscillations in a superconducting ring.
The results are shown for a niobium ring with
a radius $R=2\times 10^{-3}cm$. The cross sectional 
area of the wire (from which the ring is made from) is
$S_{0}=5\times 10^{-14}cm$. 
{\bf (a)} The frequency $\omega$
for different values of $\Phi/\Phi_0$ (the magnetic flux in
units of the flux quantum), as a function of the mode
number $\nu$.
{\bf (b)} The frequency $\omega$
as a function of the dimensionless flux, $\Phi/\Phi_0$, plotted
for different values of the wave vector $\nu$, illustrating
oscillatory behavior.
}

\newpage

%%% FIGURE 1
\begin{figure}[h!]
\begin{center}
\includegraphics[width=0.45\textwidth]{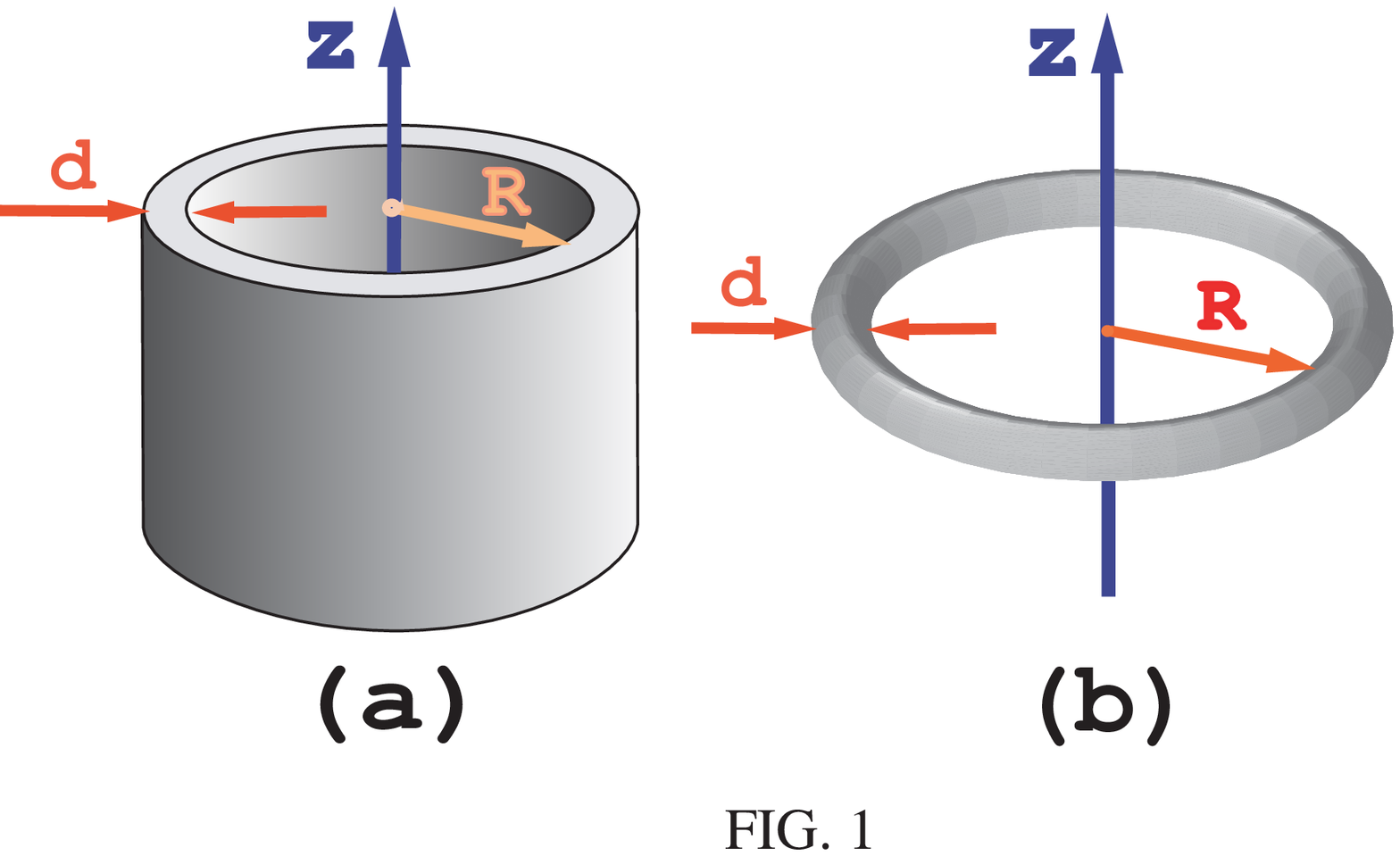}
\end{center}
\caption{}
\label{sc-cylinder-and-ring}
\end{figure}

\newpage

%%% FIGURE 2
\begin{figure}[h!]
\begin{center}
\includegraphics[width=0.48\textwidth]{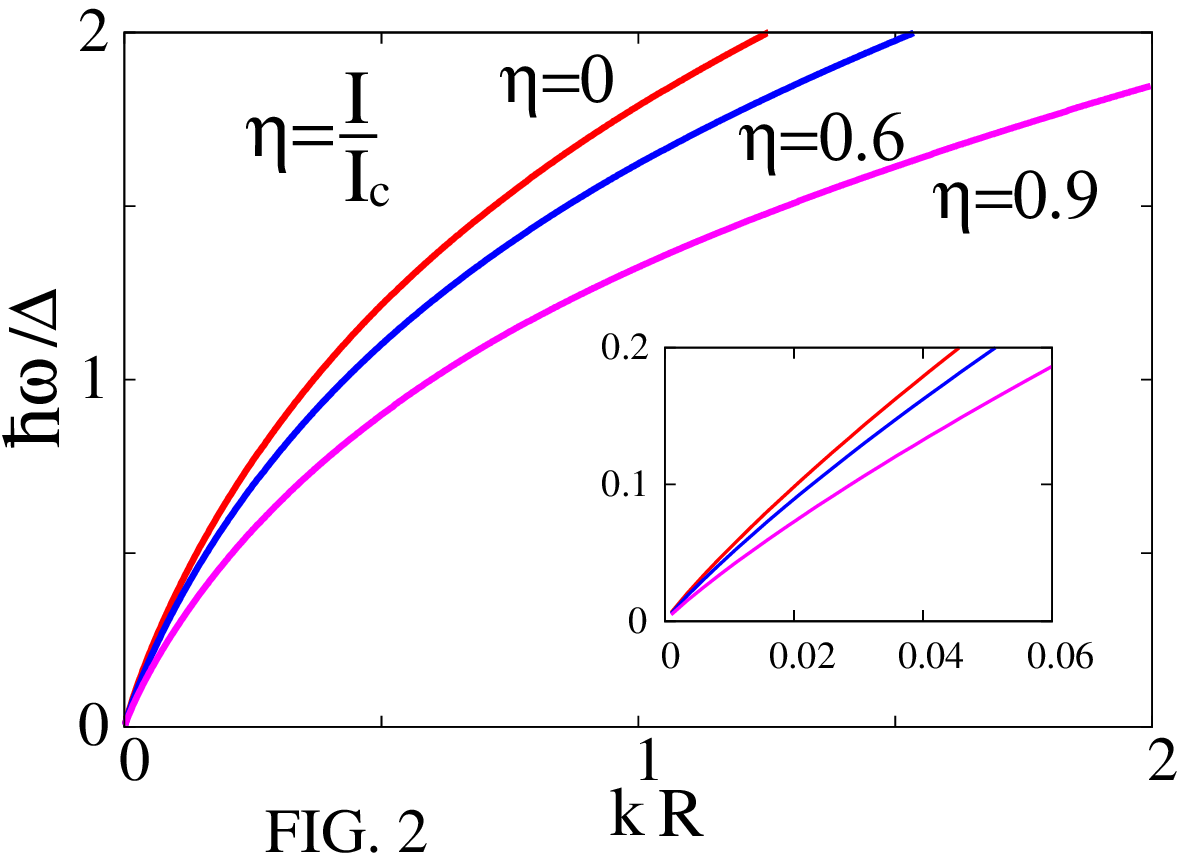}
\end{center}
\caption{}
\label{om_vs_kz_cylinder3}
\end{figure}

\newpage

%%% FIGURE 3
\begin{figure}[h!]
\begin{center}
\includegraphics[width=0.48\textwidth]{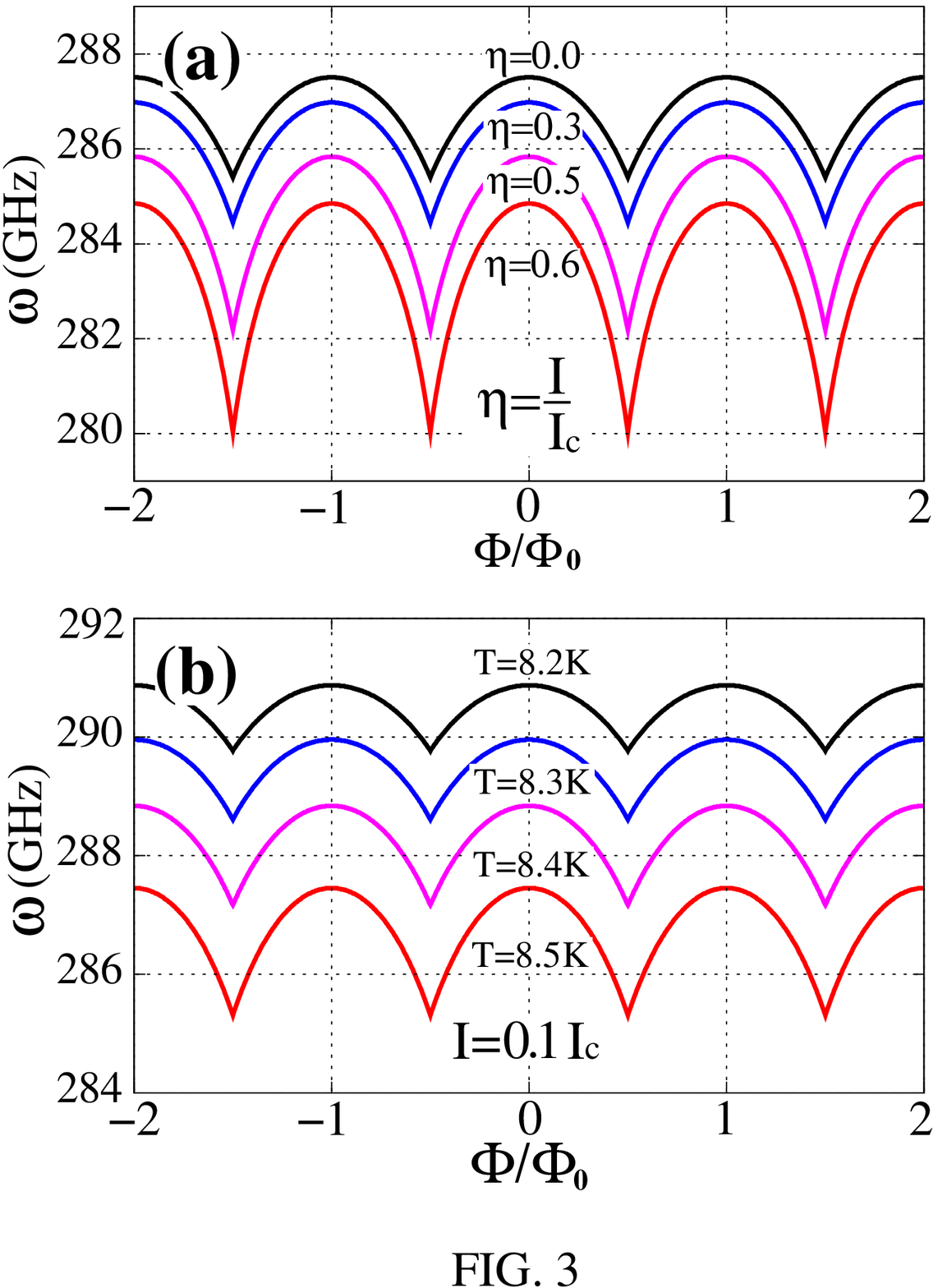}
\end{center}
\caption{}
\label{tube-plasmons}
\end{figure}

\newpage

%%% FIGURE 4
\begin{figure}[h!]
\begin{center}
\includegraphics[width=0.48\textwidth]{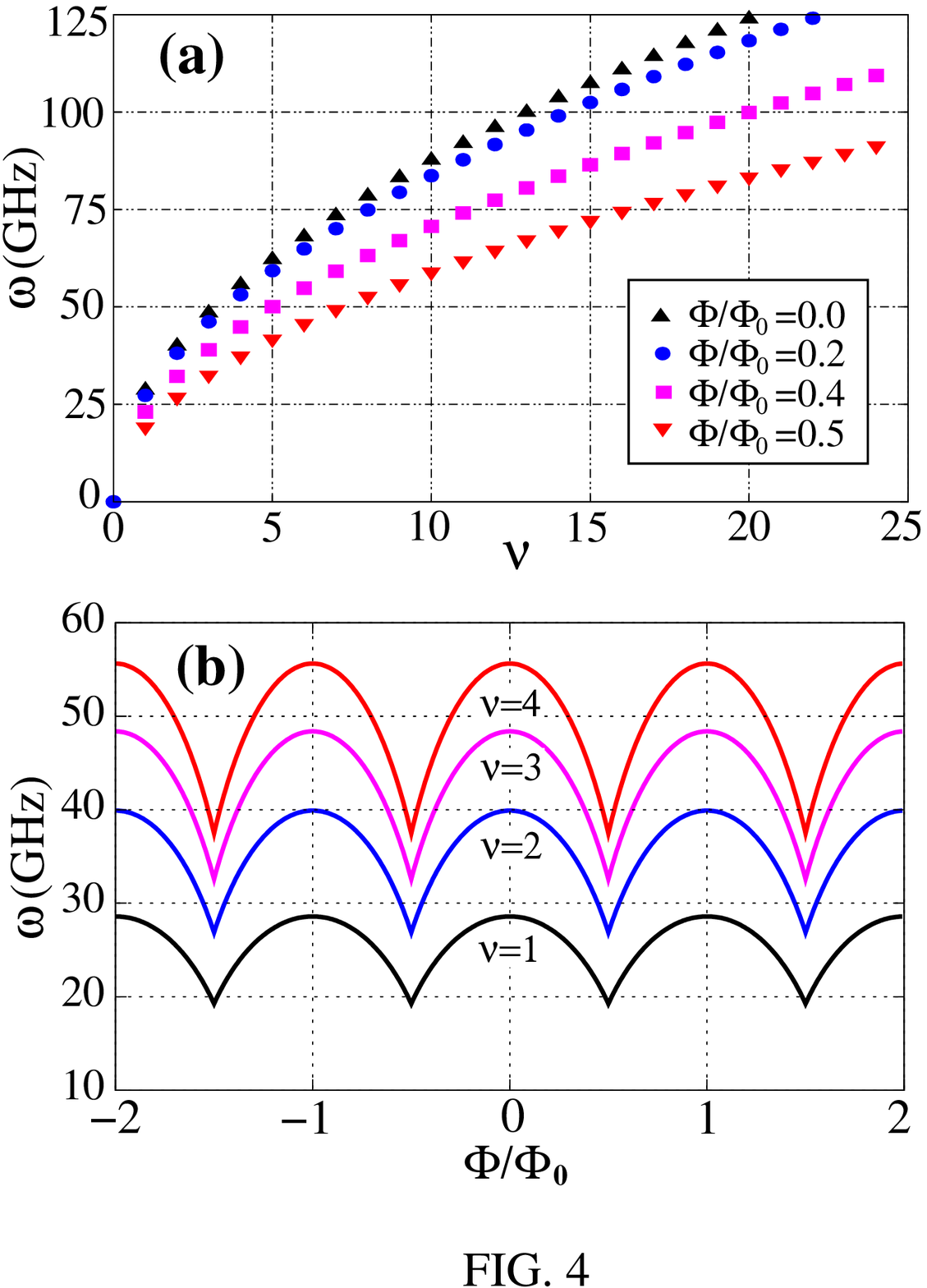}
\end{center}
\caption{}
\label{ring-plasmons}
\end{figure}


\begin{references}
\bibitem{CG_ref}
Since the phase velocity of the 
Carlson-Goldman  (CG) mode is much smaller than the phase
velocity of the acoustic plasmon mode, these two modes 
occur at very different wave vector ranges. 
Although both the acoustic and the CG type of excitations may 
exist in superconducting tubes and rings simultaneously, 
the CG mode has much smaller wavelengths. 
Because of the different nature of these modes the interaction 
between them can be neglected. Consideration of 
the influence of the Little-Parks effect on the CG mode in tubes 
and rings is left for a future study.
\bibitem{note2}
We follow the approach that was used in Ref. 5  for studying
plasma waves in thin superconducting thin wires and films.
\end{references}
\end{document}